\documentclass[12pt]{article}
\usepackage[english]{babel}
\usepackage[left=1in,right=1in,top=1in,bottom=1in]{geometry}

\usepackage[toc,page]{appendix}
\usepackage{amsmath,amssymb}
\usepackage{graphicx}
\usepackage{epstopdf}
\usepackage{float}

\usepackage{hyperref}
\usepackage{authblk}

\hypersetup{colorlinks=true, linkcolor=blue, citecolor=red, bookmarks=false, pdfstartview={FitH}}

\begin{document}

\title{\bf Partial wave analysis of the Dirac fermions scattered from Reissner-Nordstr\" om charged black holes}

\author[1]{\bf Ion I. Cotaescu\thanks{cota@physics.uvt.ro}}
\author[1]{\bf Cosmin Crucean\thanks{crucean@physics.uvt.ro}}
\author[1]{\bf Ciprian A. Sporea\thanks{ciprian.sporea89@e-uvt.ro}}
\affil[1]{\small Faculty of Physics, West University of Timi\c soara, V.  P\^ arvan Ave.  4, 300223 Timi\c soara, Romania}

\date{\small \today}
\maketitle

\begin{abstract}
The asymptotic form of Dirac spinors in the field of the Reissner-Nordstrom black hole are derived for the scattering states (with $E>mc^2$) obtaining the phase shifts of the partial wave analysis of Dirac fermions scattered from charged black holes. The elastic scattering and absorption are studied giving analytic formulas for the partial amplitudes and cross sections. A  graphical study is performed for analysing  the differential cross section (forward/backward scattering) and the polarization degree as functions of scattering angle.

\ \


\end{abstract}

\newpage

\section{Introduction}

The  problem  of the quantum fermions scattered from Schwarzschild black holes was studied either in particular cases \cite{FHM,FHM1} or by using combined  analytical and numerical methods  \cite{S1}-\cite{S3}. Recently we performed an analytic study of this process proposing a version of partial wave analysis that allowed us to write down closed formulas for the scattering amplitudes and cross sections \cite{CCS}. The analytic approach improves our understanding of the quantum mechanisms that governs the fermion scattering by black holes.

In the present letter we would like to extend this analytic study to the problem of the Dirac fermions scattered from Reissner-Nordstr\" om  charged black holes since it seems that this problem was neglected so far. The studies performed in the existing literature have been concentrating manly on scalar field \cite{bh3,bh4,RN1,RN3} and electromagnetic scattering \cite{bh5,RN2,RN2a,RN5}  on charged black holes . For these reasons we concentrate in studding the problem of fermion scattering on a Reissner-Nordstr\" om  charged black hole by using analytically and graphically methods. This phenomenon could be interesting since along side with the gravitational interaction we can study the effect of the interaction between the charges of the black hole and the fermion charge upon the scattering process. In addition the results related to the fermion absorbtion by the Schwarzschild black hole will be modified depending on the electric atraction/repulsion between the black hole charges and the fermion charge.

We deduce the asymptotic form of the Dirac spinors in the Reissner-Nordstr\" om  geometry deriving the phase shifts and the partial amplitudes of elastic scattering as well as the absorption cross section. We present the principal analytic results without extended examples or comments that exceed the space of this short paper.
We use the methods and notations of Ref. \cite{CCS} and Planck's natural units with $G=c=\hbar=1$.

\section{Asymptotic  spinors in Reissner-Nordstr\" om  geometry}

The Dirac equation in curved spacetimes  is defined in frames $\{x;e\}$  formed by a local chart of coordinates $x^{\mu}$, labeled by natural indices, $\alpha,..,\mu, \nu,...=0,1,2,3$, and an orthogonal  local frame and coframe  defined by the gauge  fields
(or tetrads), $e_{\hat\alpha}$ and  respectively $\hat e^{\hat\alpha}$, labeled by the local indices $\hat\alpha,..,\hat\mu,...$ with the same range. In local-Minkowskian manifolds $(M,g)$, having as flat model the Minkowski spacetime $(M_0,\eta)$ of metric $\eta={\rm
diag}(1,-1,-1,-1)$, the gauge fields satisfy the  usual duality
conditions, $\hat e^{\hat\mu}_{\alpha}\,
e_{\hat\nu}^{\alpha}=\delta^{\hat\mu}_{\hat\nu},\,\, \hat
e^{\hat\mu}_{\alpha}\, e_{\hat\mu}^{\beta}=\delta^{\beta}_{\alpha}$
and the orthogonality relations, $e_{\hat\mu}\cdot
e_{\hat\nu}=\eta_{\hat\mu \hat\nu}\,,\, \hat e^{\hat\mu}\cdot \hat
e^{\hat\nu}=\eta^{\hat\mu \hat\nu}$. The gauge fields define the
1-forms $\omega^{\hat\mu}=\hat e^{\hat\mu}_{\nu}dx^{\nu}$ giving the
line element
$ds^2=\eta_{\hat\alpha\hat\beta}\omega^{\hat\alpha}\omega^{\hat\beta}=g_{\mu\nu}dx^{\mu}dx^{\nu}$.

Let us consider the Dirac equation, $ i\gamma^{\hat\alpha}D_{\hat\alpha}\psi - m\psi=0$
of a free spinor field $\psi$  of  mass $m$, written with our previous notations \cite{CCS} in the frame $\{x;e\}$ defined by the Cartesian gauge \cite{C1,ES},
\begin{eqnarray}
\omega^0&=&w(r)dt \,,\\
\omega^1&=&\frac{1}{w(r)}\sin\theta\cos\phi \,dr+ r\cos\theta\cos\phi \,d\theta\nonumber\\
&&- r \sin\theta\sin\phi \,d\phi\,, \\
\omega^2&=&\frac{1}{w(r)}\sin\theta\sin\phi \,dr+ r\cos\theta\sin\phi \,d\theta\nonumber\\
&&+ r \sin\theta\cos\phi \,d\phi\,, \\
\omega^3&=&\frac{1}{w(r)}\cos\theta \,dr- r \sin\theta \,d\theta\,,
\end{eqnarray}
in the central gravitational and Coulomb  field of a charged black hole of mass $M$ and charge $Q>0$ with  the Reissner-Nordstr\" om line element
\begin{equation}\label{(le)}
ds^{2}=\eta_{\hat\alpha\hat\beta}\omega^{\hat\alpha}\omega^{\hat\beta}=w(r)^2dt^{2}-\frac{dr^{2}}{w(r)^2}- r^{2} (d\theta^{2}+\sin^{2}\theta~d\phi^{2})\,,
\end{equation}
defined on the radial domain $D_{r}= (r_{+}, \infty)$ where
\begin{equation}
w(r)=\left[1-\frac{2M}{r}+\frac{Q^2}{r^2}\right]^{\frac{1}{2}}=\left(1-\frac{r_+}{r}\right)^{\frac{1}{2}}\left(1-\frac{r_-}{r}\right)^{\frac{1}{2}}\,,
\end{equation}
and $r_{\pm}=M\pm\sqrt{M^2-Q^2}$ provided $Q<M$.  The corresponding Coulomb potential gives the potential energy
\begin{equation}
\frac{Q}{r} \to V(r)=\frac{e Q}{r}\,,
\end{equation}
of the fermion carrying the elementary electric charge $e=\pm\sqrt{\alpha}$ \footnote{In this system $\alpha\simeq \frac{1}{137}$ is the fine structure constant while  the electron mass is $m_e=\sqrt{\alpha_G}\simeq 4.178\, 10^{-23}$.}. In what follows we study the scattering solutions of the Dirac equation in the asymptotic domain where $r\gg r_+$.

We have shown \cite{C1} that in the gauge we consider here the spherical variables of the Dirac equation can be separated just as in the case of the central problems in Minkowski  spacetime \cite{TH}. Consequently, the particle-like energy eigenspinors  of energy $E$,
\begin{eqnarray}
&&U_{E,\kappa,m_{j}}({x})=U_{E,\kappa,m_{j}}(t,r,\theta,\phi)=\frac{e^{-iEt}}{rw(r)^{\frac{1}{2}}}\nonumber\\
&&\times[f^{+}_{E,\kappa}(r)\Phi^{+}_{m_{j},\kappa}(\theta,\phi)
+f^{-}_{E,\kappa}(r)\Phi^{-}_{m_{j},\kappa}(\theta,\phi)]\,,\label{(u)}
\end{eqnarray}
are expressed in terms of radial wave functions,  $f^{\pm}_{E,\kappa}$, and  usual four-component angular spinors $\Phi^{\pm}_{m_{j}, \kappa}$ \cite{TH}.  These spinors are orthogonal to each other being  labeled by the angular quantum numbers $ m_{j}$ and
\begin{equation}\label{kjl}
\kappa=\left\{\begin{array}{lcc}
~~~~\,j+\frac{1}{2}=l&{\rm for}& j=l-\frac{1}{2}\\
-(j+\frac{1}{2})=-l-1&{\rm for}& j=l+\frac{1}{2}
\end{array}\right.
\end{equation}
which encapsulates the information about the quantum numbers $l$ and
$j=l\pm\frac{1}{2}$ as defined in Refs. \cite{TH, LL} (while  in Ref.
\cite{S3}  $\kappa$ is of opposite sign).    We note that the antiparticle-like energy eigenspinors  can be obtained directly using the charge conjugation  as in the flat case \cite{C3}. Thus the problem of the angular motion is completely solved  remaining with  a pair of  radial wave functions, $f^{\pm}$, (denoted from now on without indices) which satisfy two radial equations  that can be written in compact form as the eigenvalue problem $H_r{\cal F}=E{\cal F}$ of the radial Hamiltonian \cite{C1},
\begin{equation}\label{HR}
H_r=\left(\begin{array}{cc}
    m\,w(r)+V(r)& -w(r)^2\frac{\textstyle d}{\textstyle dr}+\frac{\textstyle \kappa}
{\textstyle r}\,w(r)\\
&\\
  w(r)^2\frac{\textstyle d}{\textstyle dr}+\frac{\textstyle \kappa}
{\textstyle r}w(r)& -m\,w(r)+V(r)
\end{array}\right)\,\,,
\end{equation}
in the space of  two-component vectors,  ${\cal
F}=(f^{+}, f^{-})^{T}$,  equipped with the radial scalar product
\cite{C1}
\begin{equation}\label{(spf)}
({\cal F},{\cal F}')=\langle U,U'\rangle=\int_{D_{r}}\frac{dr}{w(r)^2}\, {\cal
F}^{\dagger}{\cal F}'\,.
\end{equation}

The resulting radial problem  cannot be solved analytically as it stays such that we are forced  to resort to the same method of approximation as in  Ref. \cite{CCS} by using a convenient Novikov's  dimensionless coordinate \cite{Nov,GRAV}. In the present case  we chose the Novikov coordinate corresponding to the event horizon of radius $r_+$ defined as
\begin{equation}\label{x}
x=\sqrt{\frac{r}{r_{+}}-1}\,\in\,(0,\infty)\,.
\end{equation}
Then, by changing the variable, multiplying with $x^{-1}(1+x^2)$ and introducing the notations
\begin{equation}
\mu=r_{+}m\,,\quad \varepsilon=r_{+}E\,,\quad \delta=\sqrt{\frac{r_-}{r_+}}\,,
\end{equation}
we rewrite the exact radial problem in the form ${\cal E} {\cal F}=0$ where the new matrix operator (\ref{rrr})
\begin{equation}\label{rrr}
{\cal E}=\left(\begin{array}{cc}
 \mu\sqrt{1+x^{2}-\delta^2}-\varepsilon\left(x+\frac{\textstyle 1}{\textstyle x}\right)+\frac{\textstyle eQ}{\textstyle x}
& -\frac{\textstyle 1}{\textstyle 2}\frac{\textstyle 1+x^2-\delta^2}{\textstyle 1+x^2}\frac{\textstyle d}{\textstyle
dx}+
\frac{\textstyle \kappa\sqrt{1+x^2-\delta^2}}{\textstyle{1+x^2}}\\
&\\
    \frac{\textstyle 1}{\textstyle 2} \frac{\textstyle 1+x^2-\delta^2}{\textstyle 1+x^2}  \frac{\textstyle d}{\textstyle dx}+
\frac{\textstyle \kappa\sqrt{1+x^2-\delta^2}}{\textstyle{1+x^2}} & -
    \mu\sqrt{1+x^{2}-\delta^2}-\varepsilon\left(x+\frac{\textstyle 1}{\textstyle
x}\right)+\frac{\textstyle eQ}{\textstyle x}
\end{array}\right)\,,
\end{equation}
is suitable for further approximations.

For very large values of $x$, we can use the Taylor expansion of these equations with respect to $\frac{1}{x}$  neglecting  the terms of the order $O(1/x^2)$. We obtain thus  the {\em asymptotic} radial problem \cite{C4,CCS} which can be rewritten as ${\cal E}'{\cal F}=0$ where the new matrix operator takes the form
\begin{equation}\label{RA}
{\cal E}'=\left(\begin{array}{cc}
 \frac{\textstyle 1}{\textstyle 2} \frac{\textstyle d}{\textstyle dx}
 +\frac{\textstyle\kappa}{\textstyle x}
& -x(\mu+\varepsilon)- \frac{\textstyle 1}{\textstyle x}(\zeta+\beta)\\
&\\
x(\varepsilon -\mu)-\frac{\textstyle 1}{\textstyle x}(\zeta-\beta)
&
   \frac{\textstyle 1}{\textstyle 2} \frac{\textstyle d}{\textstyle dx}
   -\frac{\textstyle \kappa}{\textstyle x}
\end{array}\right)\,,
\end{equation}
after reversing between themselves its lines and introducing the  notations
\begin{eqnarray}
\zeta&=&\frac{m}{2}(r_+-r_-)=\frac{1}{2}\,\mu(1-\delta^2)\,, \\
 \beta&=& \varepsilon -e Q=\varepsilon-e\delta r_+\,.
\end{eqnarray}
As in the Dirac-Coulomb case \cite{LL} it is useful to put in diagonal form the terms
 proportional to $x$ by using the matrix
\begin{equation}
T=\left(
\begin{array}{cc}
-i\sqrt{\mu+\varepsilon}&i\sqrt{\mu+\varepsilon}\\
\sqrt{\varepsilon-\mu}&\sqrt{\varepsilon-\mu}\\
\end{array}\right)\,,
\end{equation}
for transforming the radial doublet as ${\cal F}\to \hat{\cal F}=T^{-1}{\cal F}=(\hat
f^{+},\,\hat f^{-})^T$. Then we obtain the new system of radial equations
\begin{equation}\label{TE}
\frac{x}{2}
\frac{d\hat f^{\pm}}{dx}\pm i \left(\frac{\zeta\mu- \beta\varepsilon}{\nu}-\nu x^2
\right)\hat f^{\pm}
=\left(\kappa\pm i\frac{\zeta\varepsilon-\beta\mu}{\nu}\right)\hat
f^{\mp}\,,
\end{equation}
where  $\nu=\sqrt{\varepsilon^2-\mu^2}$. These equations can be  solved
analytically  for any values of $\varepsilon$ but here we restrict ourselves to the scattering modes corresponding to the continuous spectrum  $\varepsilon\in [\mu, \infty)$. These solutions can be expressed  in terms of Whittaker functions as \cite{C4,CCS}
\begin{eqnarray}
\hat f^+(x)&=&C_1^+\frac{1}{x}M_{\rho_+,s}(2i\nu x^2)
+C_2^+\frac{1}{x}W_{\rho_+,s}(2i\nu x^2)\,,\label{E11}\\
\hat f^-(x)&=&C_1^-\frac{1}{x}M_{\rho_-,s}(2i\nu x^2)
+C_2^-\frac{1}{x}W_{\rho_-,s}(2i\nu x^2)\,,\label{E22}
\end{eqnarray}
where  we denote
\begin{equation}
s=\sqrt{\kappa^2+\zeta^2-\beta^2},\quad
\rho_{\pm}=\mp\frac{1}{2}-i q,\quad q= \frac{\beta\varepsilon-\zeta\mu}{\nu} \,.
\end{equation}
The  integration constants must satisfy \cite{C4}
\begin{equation}\label{C1C1}
\frac{C_1^-}{C_1^+}=\frac{s-i q}{\kappa-i\lambda}\,,\quad
\frac{C_2^-}{C_2^+}=-\frac{1}{\kappa-i\lambda}\,,\quad \lambda=\frac{\beta\mu-\zeta\varepsilon}{\nu}\,.
\end{equation}
We observe that the functions
$M_{\rho_{\pm},s}(2i\nu x^2)=(2i\nu x^2)^{s+\frac{1}{2}}[1+O(x^2)]$ are
regular in $x= 0$, where the functions $W_{\rho_{\pm},s}(2i\nu x^2)$
diverge as $x^{1-2s}$ if $s>\frac{1}{2}$ \cite{NIST}. These
solutions will help us to find the scattering amplitudes of the
Dirac particles by charged black holes, after fixing the integration
constants.

\section{Partial wave analysis}

The scattering of Dirac fermions on a charged black hole   is described by the energy eigenspinor $U$ whose asymptotic form,
\begin{equation}
U\to U_{plane}(\vec{p})+A(\vec{p},\vec{n}) U_{sph}\,,
\end{equation}
for $r\to \infty$  (where the gravitational and Coulomb fields vanish) is
given by the plane wave spinor of momentum $\vec{p}$ and the free
spherical spinors of the flat case behaving as
\begin{equation}\label{Up}
U_{sph}\propto \frac{1}{r}\,e^{ipr-iEt}\,,\quad p=\sqrt{E^2-m^2}=\frac{\nu}{r_+}\,,
\end{equation}
since in the asymptotic zone the fermion energy is that of special relativity, $E=\sqrt{m^2+p^2}$. Here we fix the geometry such that $\vec{p}=p\vec{e_3}$ while the
direction  of the scattered fermion is given by the scattering
angles $\theta$ and $\phi$ which are just the spheric angles of the
unit vector $\vec{n}$.  Then the scattering amplitude
\begin{equation}\label{Ampl}
A(\vec{p},\vec{n})=f(\theta)+ig(\theta)\frac{ \vec{p}\land \vec{n}}{|\vec{p}\land \vec{n}|}\cdot\vec{\sigma}
\end{equation}
depend on  two scalar amplitudes, $f(\theta)$ and $g(\theta)$, that
can be  studied by using the partial wave analysis.

\subsection{Phase shifts}

The partial wave analysis exploits the asymptotic form of the exact
analytic  solutions which satisfy suitable boundary conditions that
in our case might be fixed at the (exterior) event horizon (where $x=0$).
Unfortunately, here we have only the asymptotic solutions
(\ref{E11}) and (\ref{E22}) whose integration constants cannot be
related to those of the solutions near event horizon without resorting to numerical methods \cite{FHM}-\cite{S3}. Therefore, we must chose suitable  asymptotic conditions for determining the integration constants. The arguments of Ref. \cite{CCS} (appendix C) show that in our approach it is necessary to adopt the  general  asymptotic condition  $C_2^+=C_2^-=0$  in order to have {\em elastic} collisions with a correct Newtonian limit for large angular momenta.

The asymptotic form  of the doublet ${\cal F} = T\hat{\cal F}$ can be obtained as in Ref. \cite{CCS} observing that now we must replace $\nu x^2=p(r-r_+)$. Thus we obtain  the
definitive asymptotic form of the radial functions of the scattered fermions  by charged black holes,
\begin{eqnarray}
{\cal F}&=&\left(
\begin{array}{c}i\sqrt{\varepsilon+\mu}\,(\hat f^- -\hat f^+)\\
\sqrt{\varepsilon-\mu}\,(\hat f^+ +\hat f^-)
\end{array}\right)\nonumber\\
&\propto& \begin{array}{c}
\sqrt{E+m}\,\sin\\
\sqrt{E-m}\,\cos
\end{array}\left( pr-\frac{\pi l}{2} +\delta_{\kappa}+\vartheta(r)\right)\,,
\end{eqnarray}
whose point-independent phase shifts $\delta_{\kappa}$  give the quantities
\begin{equation}\label{final}
S_{\kappa}=e^{2i\delta_{\kappa}}=\left(\frac{\kappa-i\lambda}{s-iq}\right)\,\frac{\Gamma(1+s-iq)}{\Gamma(1+s+iq)} e^{i\pi(l-s)}\,.
\end{equation}
Notice that the values of  $\kappa$ and $l$ are related as in Eq. (\ref{kjl}), i. e. $l=|\kappa|-\frac{1}{2}(1-{\rm sign}\,\kappa)$. The remaining point-dependent phase,
\begin{equation}
\vartheta(r)= -p r_++q \ln [2p(r-r_+)]\,,
\end{equation}
which does not depend on angular quantum numbers, may be ignored as
in the  Dirac-Coulomb case \cite{LL,S3}.

We arrived thus at the final result (\ref{final}) depending on the parameters  introduced above that can be expressed in terms of physical quantities $m,\,M,\, e,\,,Q...$ etc..
In addition, assuming that the inequality $|\beta|\ge|\zeta|$ holds even for $p=0$,  it  is convenient to introduce the new real parameter $k$ obeying
\begin{equation}\label{kqlam}
k^2=\beta^2-\zeta^2=q^2-\lambda^2\,,
\end{equation}
that allows us to write simply $s=\sqrt{\kappa^2-k^2}$. Then by using Eq. (\ref{Up}b) we obtain our principal new result that holds for massive fermions:
\begin{eqnarray}
 k&=&\left[(r_+E-eQ)^2-\frac{m^2}{4}(r_+-r_-)^2\right]^{\frac{1}{2}}\,, \label{k}\\
q&=&r_+p+M\frac{m^2}{p}-eQ\frac{E}{p}
\,,\label{q}\\
\lambda &=&mM\frac{E}{p}-eQ\frac{m}{p}\,, \label{lam}
\end{eqnarray}
Obviously, in the case of the massless neutral fermions (with $m=e=0$) we remain with  the unique parameter $q=r_+p$ since then $\lambda=0$ and  $k=q$. Hereby we see that the parameter $s$ has a special position since this can take either real  values  or pure imaginary ones regardless the fermion mass.

\subsection{Elastic scattering}

For the real values of $s$ the scattering is elastic since in this case the identity (\ref{kqlam})  guarantees that  the phase shifts of  Eq. $(\ref{final})$ are real numbers  such that $|S_{\kappa}|=1$. Obviously, this happens only when  $\kappa$  (at given $p$)  satisfies the condition
\begin{equation}\label{cond}
|\kappa|\geq \tilde k+1\,,
\end{equation}
where $\tilde k={\rm floor} (k)$ is the greater integer less than $k$.  Then, the scalar amplitudes of Eq. (\ref{Ampl}),
\begin{eqnarray}
f(\theta)=\sum_{l=0}^{\infty}a_l\,P_l(\cos \theta)\,,\label{f} \qquad  g(\theta)=\sum_{l=1}^{\infty}b_l\,P_l^1(\cos\theta)\,, \label{g}
\end{eqnarray}
which depend on the following {\em partial} amplitudes \cite{LL,S3},
\begin{eqnarray}
a_l&=&(2l+1)f_l\nonumber=\frac{1}{2ip}\left[(l+1)(S_{-l-1}-1)+l(S_l-1)\right]\,,\label{fl}\\
b_l&=&\!\!\!(2l+1)g_l=\frac{1}{2ip}\left(S_{-l-1}-S_l\right)\,,\label{gl}
\end{eqnarray}
give rise to  the {\em elastic} scattering intensity or  differential cross section,
\begin{equation}\label{int}
{\cal I}(\theta)=\frac{d\sigma}{d\Omega}=|f(\theta)|^2+|g(\theta)|^2\,,
\end{equation}
and the polarization degree,
\begin{equation}\label{pol}
{\cal P}(\theta)=-i\frac{f(\theta)^*g(\theta)-f(\theta) g(\theta)^*}{|f(\theta)|^2+|g(\theta)|^2}.
\end{equation}
This last quantity is interesting for the scattering of massive
fermions  representing the induced polarization for an unpolarized
initial beam.

\subsection{Absorption}

The absorption is present in the partial waves for which  we have
\begin{equation}\label{cond1}
1\le |\kappa|\leq \tilde k\,.
\end{equation}
Here we meet a branch point in $s=0$ and two solutions $s=\pm
i|s|=\pm i\sqrt{k^2-\kappa^2}$,  among them we must chose $s=-i|s|$
since only in this manner we select the physical case of
$|S_{\kappa}|<1$. More specific, by substituting $s=-i|s|$ in Eq.
(\ref{final})  we obtain the simple closed form
\begin{equation}\label{SSk}
|S_{\kappa}|=|S_{-\kappa}|=e^{-2\Im \delta_{\kappa}}=e^{-\pi |s|}\sqrt{\frac{\sinh \pi (q-|s|)}{\sinh \pi (q+|s|)}}
\end{equation}
showing that $0< |S_{\kappa}|< 1$ since $|s|<q$ for any $(p,\kappa)$
obeying the condition (\ref{cond1}).  Moreover, we can verify that
in the limit of the large momentum (or energy) the absorption tends
to become maximal since
\begin{equation}\label{limS}
\lim_{p\to \infty} |S_{\kappa}|=0\,,
\end{equation}
regardless of the fermion mass.

Under such circumstances we can calculate the  absorption cross section that reads
 \cite{S3}
\begin{equation}\label{sigmaaa}
\sigma_a=\sum_{l\ge 1}\sigma_a^l (p)=\frac{2\pi}{p^2}\sum_{l=1}^{\tilde k}l (1-|S_l|^2)\,,
\end{equation}
since for $s=-i|s|$ we have $|S_{-\kappa}|=|S_{\kappa}|$ as in Eq. (\ref{SSk}).
This  cross section can be calculated at any time as a finite sum  of the partial cross sections whose  definitive closed form,
\begin{eqnarray}\label{sigmaa}
&&\sigma_a^l(p)=\theta(k-l)\frac{2\pi l}{p^2}\nonumber\\
&&~~~\times \left[1-e^{-2\pi\sqrt{k^2-l^2}}\, \frac{\sinh
\pi(q-\sqrt{k^2-l^2})}{\sinh \pi(q+\sqrt{k^2-l^2})}\right]\,,
\end{eqnarray}
is derived according to Eqs. (\ref{SSk}) and (\ref{sigmaaa}) while
the condition (\ref{cond1})  introduces the Heaviside step function
$\theta(k-l)$. Hereby we understand that the absorption arises in
the partial wave $l$ for the values of $p$ (or $E$) satisfying the condition
$k>l$. This means that for any fixed value of $l$  there is a threshold, $E_l$, defined as the positive solution of the equation
\begin{equation}\label{condx}
|r_+E_l-eQ|=\sqrt{l^2+\frac{1}{4}m^2(r_+-r_-)^2}\,.
\end{equation}
A similar condition with \ref{condx} was obtained in ref. \cite{abs} for absorbtion on a dilaton black hole.
This indicates that the fermions with $|\kappa|=l$ can be absorbed by
black hole only if $E\ge E_l$.  The existence of these thresholds is important since these keep under control the effect of the singularities in $p=0$.

Finally, we observe that in the high-energy  limit  all these absorption cross sections tend to the event horizon (apparent) area,
indifferent on the fermion mass  $m\ge 0$, as it results from  Eq.
(\ref{limS}) that yields
\begin{equation}\label{asy}
\lim_{p\to \infty}\sigma_a =\lim_{p\to
\infty}\frac{2\pi}{p^2}\,\sum_{l=1}^{\tilde k}  1=\lim_{p\to
\infty}\frac{\pi}{p^2}\,\tilde k(\tilde k+1)=\pi r_+^2
\end{equation}
since  $\tilde k(\tilde k+1)\sim k^2\sim r_+^2 p^2$.

\section{Graphical discussion of the results}

Let us now discuss some physical consequences of our results encapsulated by the formulas presented in the previous sections. For a better understanding of the analytical results we perform a graphical analysis of the differential cross section in terms of scattering angle $\theta$. All the plots are obtained using the methods described in Ref.\cite{CCS} where we used a technique proposed some time ago by Yennie et al.\cite{Yeni}. In what follows we focus our analysis on scattering from small or micro black holes (with $M \sim 10^{-15}$ kg) since in this case the wave length of the fermion ($\lambda=2\pi h/p$) and the Schwarszchild radius ($r_S=2M$) have the same order of magnitude such that we can observe the presence of glory and orbiting scattering.

\begin{figure}
\centering
\includegraphics[scale=0.6]{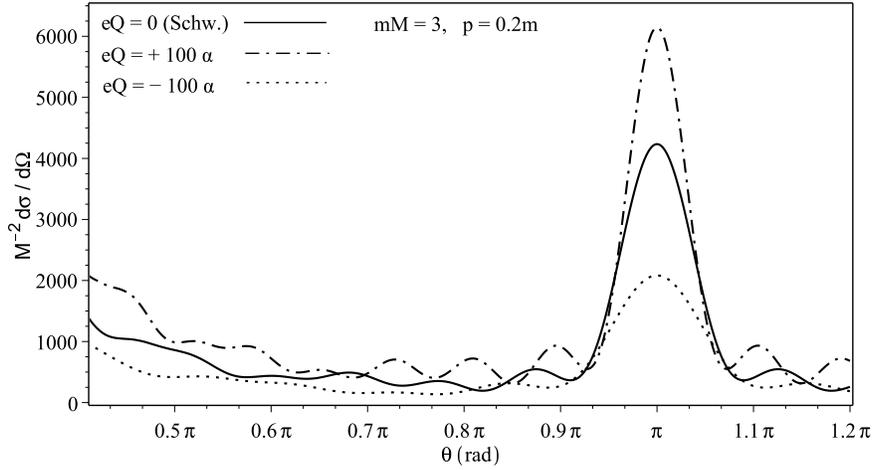}
\caption{Comparison between the scattering cross section of Reissner-Nordstr\" om black hole and the Schwarzschild black hole for $p=0.2m$ and $mM=3$.}
\label{fig.1}
\end{figure}

Comparing the scattering cross section of a Reissner-Nordstr\" om black hole with that of  a Schwarzschild black hole (see fig. \ref{fig.1}) one can observe several things. First is that in the case of scattering by the Reissner-Nordstr\" om black hole the differential cross section will increase if the charge of the black hole and the incoming fermions charge have both the same sign (the point and dash-dotted lines in fig. \ref{fig.1}). Thus we can say that the electric repulsion increases the scattering cross section as expected, according to (\ref{condx}), comparatively with the scattering on Schwarzschild black hole (the solid line in fig. \ref{fig.1}). In the opposite case, when the black hole charge and the incoming fermions charge have opposite signs then the electric attraction between the two charges will make more fermions to be absorbed by the black hole lowering thus the scattering cross section comparatively with the scattering on Schwarzschild black hole (as can be seen from the doted line in fig.\ref{fig.1}). Figure \ref{fig.1} also revel us the fact that the width of the glory peak becomes more pronounced when the sign of the total black hole charge is the same with the charge of the fermion, respectively the glory peak's width decreases if the black hole and the fermions have opposite signs. We can conclude that when the charge of the black hole and fermions charge have the same sign, the glory scattering will be a phenomenon which becomes important comparatively with the Schwarzschild case.

\begin{figure}
\centering
\includegraphics[scale=0.6]{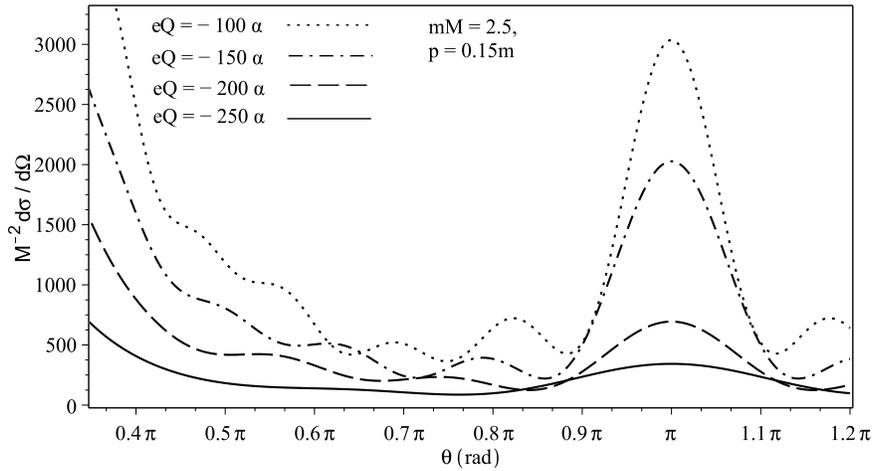}
\caption{Reissner-Nordstr\" om black hole scattering cross section for $p=0.15m, mM=2.5$ and different negative values of $eQ$. In all plots $\alpha\approx1/137$ represents the fine structure constant.}
\label{fig.2}
\end{figure}

\begin{figure}
\centering
\includegraphics[scale=0.6]{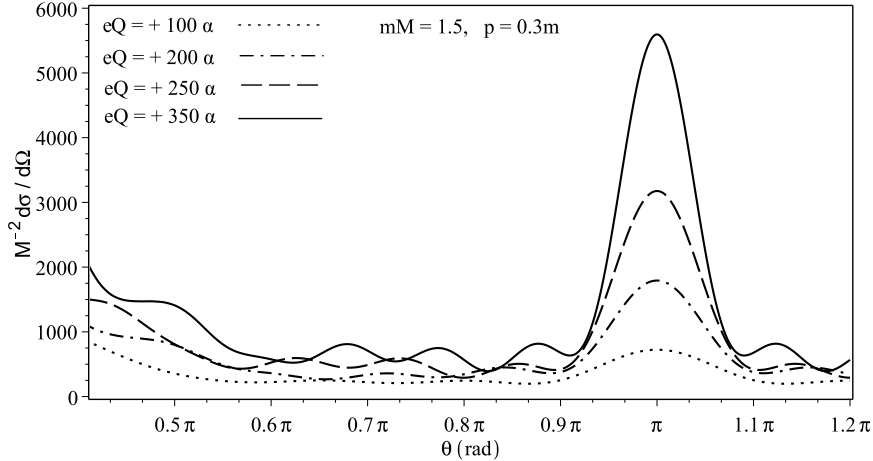}
\caption{Reissner-Nordstr\" om black hole scattering cross section for $p=0.3m, mM=1.5$ and different positive values of $eQ$ .}
\label{fig.3}
\end{figure}

In figs. \ref{fig.2}-\ref{fig.3} we present the differential cross section in terms of scattering angle for the Reissner-Nordstr\" om black hole. The effects of electric interaction (between the black hole and fermion charges) on the glory and orbiting scattering can be also observed. We have found that if the black hole has opposite charge than the incoming fermion then, as we increase the  charge on the black hole, the glory (i.e. backward scattering at angles close to $\pi$) starts to decrees up to a point when it will disappear completely (see fig.\ref{fig.2}). The same is true for the orbiting scattering (i.e. scattering for $\theta<\pi$) for which we observe a decrees in oscillations as the black hole charge increases. This should not come as a surprise since in this case the opposite signs of the charges will increase the absorption in the black hole. On the contrary if the black hole and fermion's charges have the same sign (see fig.\ref{fig.3}) then the glory will increase as the black hole becomes more charged. The same is true for the orbiting scattering for which the oscillatory behaviour becomes more pronounced as we add more charge on the black hole. Regarding the forward scattering, we have found it to be divergent for $\theta\rightarrow 0$ as can be seen from fig. \ref{fig.2}. This is in fact an effect of the long range nature of the gravitational and electric potentials, which both behave as $1/r$.
\begin{figure}
\centering
\includegraphics[scale=0.6]{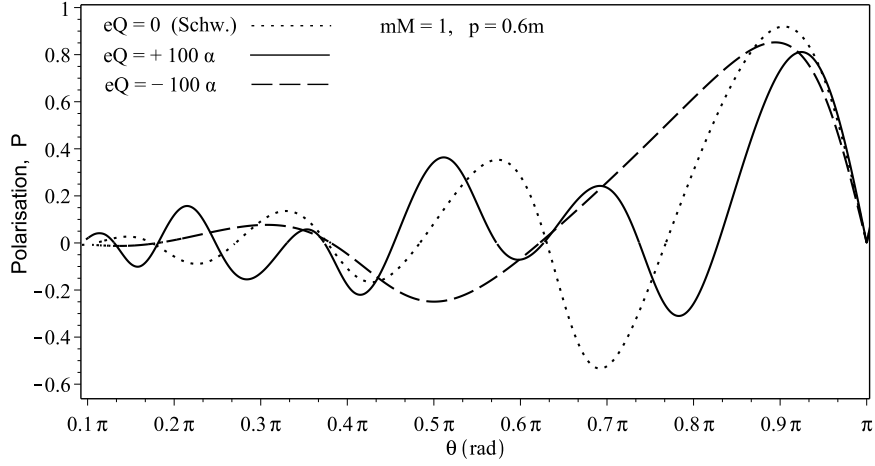}
\caption{Comparison of the Schwarzschild polarisation (doted line) with Reissner-Nordstr\" om polarisation (dashed line for positive charge on the black hole, respectively solid line for negative charge on the black hole) for $p=0.6m$ and $mM=1$.}
\label{fig.4}
\end{figure}
Regarding the effects of the black hole's charge on the induced polarisation after the scattering of an initially unpolarised beam we can say according to fig. \ref{fig.4} the followings: (i) compared with the Schwarzschild polarisation (dotted line) the Reissner-Nordstr\" om polarisation is a less oscillating function if the black hole has opposite charge that the incoming fermions (dashed line), respectively the oscillations become more fervent if the black hole charge has the same sign as the incoming fermions (solid line); (ii) the oscillations appearing in the polarisation can be seen as resulting from the oscillatory behavior of glory/orbiting scattering as well as from the  forward/backward one.
\begin{figure}
\centering
\includegraphics[scale=0.6]{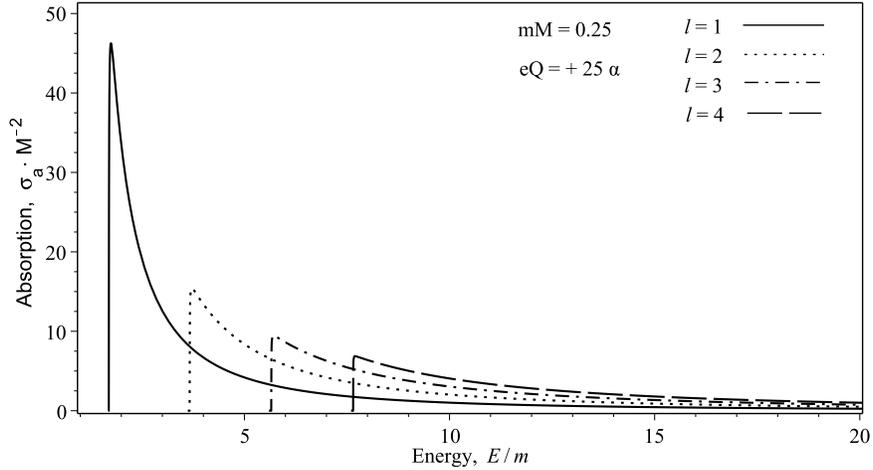}
\caption{Partial absorption cross section as a function of $E/m$ for the Reissner-Nordstr\" om black hole with $eQ=0.25\alpha$ and $mM=0.25$, for given values of angular momentum $l=1,2,3,4$.}
\label{fig.5}
\end{figure}
The dependence of the absorbtion cross section in terms of energy ($E/m$), is given in Fig.(\ref{fig.5}), where the plots were obtained for $l=1,2,3,4$. We observe that the modes with small angular momenta have the most important contribution to the absorbtion, because as we increase $l$ the maxima observed in Fig.(\ref{fig.5}) becomes smaller. Also we observe that the maxima are shifted to the right as we increase the value of $l$. Our graphical result for the absorbtion cross section is similar to those obtained in \cite{abs}, where the absorbtion of scalar particles on dilaton black hole was studied.

\section{Concluding remarks}

This is the basic framework of the relativistic partial wave
analysis of the Dirac fermions scattered by charged black
holes in which we consider exclusively the contribution of the
scattering modes. Our results are in accordance with the Newtonian
limit since in the large-$l$ limits and for very small momentum we
can take $s\sim |\kappa|\sim l$ and  $\lambda\sim q$   such that our
phase shifts (\ref{final}) becomes just the Newtonian ones \cite{FHM,FHM1}.

The above results encapsulate a new  interesting particular case  namely, the Dirac-Coulomb scattering in the presence of the gravitational field of the charged target that gives rise to a  Schwarzschild  gravitational field instead of a Reissner-Nordstr\" om one. In this situation we must take $\delta \to 0$ everywhere apart from the Coulomb term such that
$r_+\to 2M$ and $r_-\to 0$. Thus we remain with the partial waves given by Eq. (\ref{final}) but with the new parameters
\begin{eqnarray}
k&\to& \left[\left(2M E-{eQ}\right)^2-m^2M^2\right]^{\frac{1}{2}}\\
q&\to&{2Mp}+ M\frac{m^2}{p}-{eQ}\frac{E}{p}\\
\lambda&\to& m M\frac{E}{p}-eQ\frac{m}{p}\,.
\end{eqnarray}
Now we observe that for $Q=0$ we recover the results of Ref. \cite{CCS} concerning the
collision between a Dirac fermion and a Schwarzschild (neutral) black hole. Moreover, if we keep $Q\not=0$ taking $M\to 0$ we recover the well-known Dirac-Coulomb scattering in Minkowski spacetime \cite{LL}  with the parameters
\begin{equation}\label{Coul}
k\to |e Q|\,,\quad q\to -eQ\frac{E}{p}\,,\quad \lambda \to -{eQ}\frac{m}{p}\,.
\end{equation}
The  conclusion is that here we derived the most general results of the Dirac-Coulomb scattering in central gravitational fields.

\section*{Acknowledgments}

C.A. Sporea was supported by a grant of the Romanian National Authority for Scientific Research,
Programme for research-Space Technology and Advanced Research-STAR, project nr. 72/29.11.2013 between
Romanian Space Agency and West University of Timisoara.


\begin{thebibliography}{60}


\bibitem{FHM}
J. A. H. Futterman, F. A. Handler, and R. A. Matzner,
{\em Scattering from Black Holes} (Cambridge University
Press, Cambridge, England, 1988).

\bibitem{FHM1}
N. K. Kofiniti,
{\em Int. Journal Theoret. Phys.} {\bf 23}, 991 (1984).
\bibitem{S1}
J. Jing, {\em Phys. Rev. D} {\bf 70}, 065004 (2004); {\em Phys. Rev.
D} {\bf 71}, 124006 (2005).
\bibitem{S2}
K. H. C. Castello-Branco, R. A. Konoplya and A. Zhidenko, {\em Phys.
Rev. D} {\bf 71}, 047502 (2005).
\bibitem{bh1}
C. Doran, A. Lasenby, S. Dolan and I. Hinder, {\em Phys. Rev. D} {\bf 71}, 124020 (2005).
\bibitem{S3}
S. Dolan, C. Doran and A. Lasenby, {\em Phys. Rev. D} {\bf 74}, 064005 (2006).
\bibitem{CCS}
 I.I. Cot\u aescu, C. Crucean and C.A. Sporea, {\em Eur. Phys. J. C}, DOI 10.1140/epjc/s10052-016-3936-9 (2016); arXiv:1409.7201.
\bibitem{bh4}
C. L. Benone, E. S. Oliveira, S. R. Dolan and L. C. B. Crispino, {\em Phys. Rev. D} {\bf 89}, 104053 (2014).
\bibitem{RN1}
 L.C.B. Crispino, S.R. Dolan and E.S. Oliveira, {\em Phys. Rev. D}{\bf 79},064022 (2009).


\bibitem{RN3}
C.F.B. Macedo  and L.C.B. Crispino, {\em Phys. Rev. D}{\bf 90}, 064001 (2014).

\bibitem{bh3}
R. A. Konoplya and A. Zhidenko, {\em Phys. Rev. D} {\bf 76}, 084018 (2007).

\bibitem{bh5}
L. C. B. Crispino, S. R. Dolan, E. S. Oliveira, {\em Phys. Rev. Lett.} {\bf 102}, 231103 (2009).

\bibitem{RN2}
 L.C.B. Crispino, S.R. Dolan, A. Higuchi and E.S. Oliveira, {\em Phys. Rev. D}{\bf 90}, 064027 (2014).

\bibitem{RN2a}
 L.C.B. Crispino, S.R. Dolan, A. Higuchi and E.S. Oliveira, {\em Phys. Rev. D}{\bf 92}, 084056 (2015).

\bibitem{RN5}
 E.S. Oliveira, L.C.B. Crispino and A. Higuchi, {\em Phys. Rev. D}{\bf 84}, 084048 (2011).

\bibitem{C1}
I. I. Cot\u aescu, {\em Mod. Phys. Lett. A} {\bf 13}, 2923   (1998).

\bibitem{ES}
I. I. Cot\u aescu, {\em J. Phys. A: Math. Gen.} {\bf 33}, 1977
(2000).

\bibitem{TH}
B. Thaller,  {\it The Dirac Equation} (Springer Verlag, Berlin
Heidelberg, 1992).

\bibitem{LL}
V. B. Berestetski, E. M. Lifshitz and L. P. Pitaevski, {\em Quantum Electrodynamics} (Pergamon Press, Oxford 1982).

\bibitem{C3}
I. I. Cot\u aescu, {\em Phys. Rev. D} {\bf 60}, 124006 (1999).

\bibitem{Nov}
I. D. Novikov, {\em doctoral disertation}, Sthernberg Astronomical Institute (1963).


\bibitem{GRAV}
C. W. Misner, K. S. Thorne and J. A. Wheeler, {\em Gravitation} (Freeman \& Co., San Francisco, 1971).

\bibitem{C4}
I. I. Cot\u aescu, {\em Mod. Phys. Lett. A} {\bf 22}, 2493  (2007).


\bibitem{NIST}
F. W. J. Olver, D. W. Lozier, R. F. Boisvert and C. W. Clark, {\em NIST Handbook of Mathematical Functions} (Cambridge University Press, 2010).

\bibitem{abs}
I. Sakalli, O.A. Aslan, {\em Astroparticle Physics} {\bf 74}, 73–78 (2016).

\bibitem{Yeni}
D. R. Yennie, D. G. Ravenhall, and R. N. Wilson, {\em Phys. Rev.} {\bf 95}, 500 (1954).


\end{thebibliography}
\end{document}